\documentclass[twocolumn,showpacs,preprintnumbers,amssymb,pra]{revtex4}
\usepackage{pgfplots}
\usepackage{latexsym,epsfig}
\usepackage{graphicx}
\usepackage{dcolumn}
\usepackage{amsthm,amsmath}

\begin{document}

\title{Dipole Polarizability Calculation of Cd Atom: Inconsistency with experiment}

\author{B. K. Sahoo}
\email{bijaya@prl.res.in}
\affiliation{Atomic, Molecular and Optical Physics Division, Physical Research Laboratory, Navrangpura, Ahmedabad 380009, India and\\
State Key Laboratory of Magnetic Resonance and Atomic and Molecular Physics, Wuhan Institute of Physics and Mathematics,
Chinese Academy of Sciences, Wuhan 430071, China}

\author{Yan-mei Yu}
\email{ymyu@aphy.iphy.ac.cn}
\affiliation{Beijing National Laboratory for Condensed Matter Physics, Institute of Physics, Chinese Academy of Sciences, Beijing 100190, China}

\date{Received date; Accepted date}

\begin{abstract}
 Three earlier relativistic coupled-cluster (RCC) calculations of dipole polarizability ($\alpha_d$) of the Cd atom are not in good 
agreement with the available experimental value of $49.65(1.65) \ e a_0^3$. Among these two are finite-field approaches in which the 
relativistic effects have been included approximately, while the other calculation uses a four component perturbed RCC method. However, 
another work adopting an approach similar to the latter perturbed RCC method 
gives a result very close to that of experiment. The major difference between these two perturbed RCC approaches lies in their 
implementation. To resolve this ambiguity, we have developed and employed the relativistic normal coupled-cluster (RNCC) theory to 
evaluate the $\alpha_d$ value of Cd. The distinct features of the RNCC method are that the expression for the expectation  
value in this approach terminates naturally and that it satisfies the Hellmann-Feynman theorem. In addition, we determine this quantity 
in the finite-field approach in the framework of A four-component relativistic coupled-cluster theory. Considering the results from both these 
approaches, we arrive at a reliable value of $\alpha_d=46.02(50) \ e a_0^3$. We also demonstrate that the contribution from the triples 
excitations in this atom is significant.
\end{abstract}

\maketitle

\section{Introduction}

Accurate values of the electric dipole polarizabilities ($\alpha_d$) of atomic states are necessary for high precision experiments on 
optical lattices, atomic clocks, quantum information, and many other important areas of atomic and molecular physics 
\cite{manakov,bonin,pethick,madej,bks-clock}. Comparisons between the calculated $\alpha_d$ values and experimental results could serve as 
benchmarks to validate many-body methods \cite{kello,seth,yashpal1,yashpal2,sidh}. Methods that are capable of yielding results in 
close agreement with high precision experimental results are considered to be accurate and suitable for the evaluation of properties 
of atomic systems and their values can be treated as reliable when experimental results are not available. Many-body calculations 
are performed using finite-size many-electron and single-electron basis wave functions as approximations have to be made in determining 
higher order correlation effects due to limitations of computational resources. A large number of numerical operations are performed, thus 
it is not possible to estimate uncertainties in the calculations due to numerical truncations. In such a situation, just a comparison of a 
calculated value with an experimental result cannot reliably validate a method \cite{pradeep}. Therefore, it is imperative to perform 
calculations using many-body methods that can capture a wide range of physical effects and have the merit of capturing correlation effects
to all orders of the residual Coulomb interaction at different levels of approximation and are size-extensive in order to apply them for
high precision studies. To ascertain the accuracies of the results, it is necessary to check the consistencies in the results by employing
a number of theories that are equivalent to all-order many-body perturbation methods. 

 Many-body perturbation theory (MBPT) was first developed by Brueckner \cite{bruck1,bruck2,bruck3} and 
Goldstone \cite{gold}. Newer versions of this theory are now widely used to calculate atomic wave functions and properties in 
many-electron systems. Important steps to determine atomic dipole polarizabilities were taken by Dalgarno and his collaborators 
\cite{dalgarno1,dalgarno2} and Kelly \cite{kelly}. The approach adopted by Dalgarno and collaborators solves an inhomogeneous differential 
equation to obtain the first-order wave function using Rayleigh-Schroedinger perturbation theory. This approach, known as the coupled-perturbed 
Hartree-Fock (CPHF) method or random phase approximation (RPA), can predict $\alpha_d$ values very accurately in some cases, but it does not
account for a number of different classes of electron correlation effects. On the other hand, the approach adopted by Kelly using the MBPT
method pioneered by Br\"uckner and Goldstone follows a diagrammatic technique in which the contributions from different types of electron 
correlation effects can be illustrated in a transparent manner. However, it is not simple to include higher-order correlation contributions
in this approach as it treats the residual Coulomb interaction Hamiltonian and the dipole operator ($D=|{\vec D}|$) as two different 
perturbations. Another suitable approach to determine $\alpha_d$ for atomic systems is to use a finite-field method, in which the 
interaction Hamiltonian due to ${\vec D}$ with an arbitrary external electric field is added to the atomic Hamiltonian to obtain the energy
eigenvalues \cite{gready,amos,dahle}. Then, the $\alpha_d$ values are inferred from the second derivative of the energy with respect to the 
electric field. The disadvantage of this approach is that it neglects the higher-order corrections to the energies due to the electric 
field. Hence, there is a loss of numerical accuracy in the results. This approach is suitable for the molecular systems where the electron 
orbitals, described by the Cartesian coordinate system, are mixed in parities and computations are minimized by utilizing group symmetry 
identities \cite{willock,dyall}. For determining $\alpha_d$ values of atoms in this approach, one can choose a special group symmetry. 
However, it cannot still describe atomic orbitals with the same accuracy as in the spherical coordinate system. It is to be noted that it 
is possible to work with mixed parity orbitals in the spherical coordinate system, but it will be computationally more expensive.

 One of the key differences between the spherical and Cartesian coordinate systems for carrying out calculations is that the atomic 
orbitals are divided into radial and angular factors in the former case. Thus, all the physical operators are expressed using spherical 
tensors to take care of the angular momentum selection rules. The coupled-cluster (CC) theory is an all-order perturbative method and it 
is size-consistent and size-extensive for which it is  referred to as the gold standard for treating correlation effects in many-electron
systems \cite{bartlett,crawford,bishopbook}. For performing CC calculations in a spherical coordinate system using atomic orbitals with 
definite parities, the two-body interactions and the CC wave operators must be expanded in terms of multipoles \cite{lindgren}. We have 
developed different methods in the relativistic CC theory framework (RCC method) to calculate $\alpha_d$ values of atomic systems in the 
spherical coordinate system \cite{yashpal1,yashpal2,sahoo2,sahoo3}. Since the atomic orbitals in this case have definite parities, we had 
perturbed the RCC wave functions by considering $D$ as the external perturbation to  first-order. This is similar in spirit of the 
aforementioned approach by Dalgarno \cite{dalgarno1,dalgarno2} in which we obtain the solution to the inhomogeneous differential equation 
in terms of the first-order perturbed RCC wave function. In addition, our RCC method also gives contributions from various electron 
correlation effects in terms of Goldstone diagrams; similar to Kelly's approach \cite{kelly}. We have applied this method to a number of 
atomic systems to determine $\alpha_d$ values very accurately \cite{yashpal1,yashpal2,sahoo2}. In one of our works, we had obtained 
$\alpha_d=45.86(15) \ ea_0^3$ for the Cd atom \cite{yashpal2} using our RCC theory, where the corresponding experimental value has been 
reported as $49.65 \pm 1.49 \pm 0.16 \ ea_0^3$ \cite{goebel};  with the net uncertainty this value is $\alpha_d=49.65(1.65) \ ea_0^3$. In 
the same study, we had also obtained these values for other atoms belonging to homologous group of Cd in the periodic table like Zn and Hg, 
which were in very good agreement with their respective experimental results \cite{yashpal2}. In fact, our findings were also in 
agreement with the previous calculations, which were obtained by applying other variants of CC theories in the finite-field procedure. These 
calculations, however, were performed using quasi-relativistic \cite{kello} and scalar two-component Douglas-Kroll \cite{seth} Hamiltonians in 
contrast to our four-component relativistic Hamiltonian to account for the relativistic effects. Following these works, another group 
has reported $\alpha_d$ value as $49.24 \ ea_0^2$ \cite{sidh} employing a perturbative RCC method like ours \cite{yashpal2} and has referred 
to it as the perturbed RCC (PRCC) method in the singles and doubles approximation and perturbed RCC with partial triples (PRCC(T)) method when 
triples effects were included. This calculation is very close to the central value of the experimental result and is in disagreement with all the 
previous calculations. Thus, it is necessary to understand the reasons for the disagreement among these theoretical calculations and find a more 
reliable value of $\alpha_d$ of the Cd atom. Analysis of these methods reveals that there were no additional physical effects included in the PRCC 
method which could be responsible for improving the result. This means that the difference in the implementation procedures for both the 
four-component perturbative RCC methods is responsible for the discrepancies between the results.

 The RCC theories employed in Refs. \cite{yashpal1,yashpal2,sidh}  are size-extensive. In the framework of these theories, the expression for the 
energies terminate, but the expectation values corresponding to different properties do not. Recently, we have observed that the inclusion of 
higher-order non-linear terms in the non-terminating series in the evaluation of $\alpha_d$ and permanent electric dipole moment (EDM) in the $^{199}$Hg
atom influence the results significantly
\cite{sahoo1}. Therefore, it is imperative to adopt a relativistic CC method in the spherical coordinate system in which the expectation value
terminates naturally. This would be particularly relevant in the evaluation of $\alpha_d$ for Cd atom where the results of the calculations from
different methods are inconsistent and differ substantially from the measured value. In this context, the normal coupled-cluster (NCC) method 
\cite{bishopbook,arponen,bishop} would be more appropriate for the evaluation of $\alpha_d$. This method satisfies the Hellman-Feynman theorem. Moreover,
in the NCC method, the expressions for both energies and expectation values corresponding to different  physical properties terminate in a natural way.
The normalization factor in this method is equal to unity. The additional effort of implementing this method for determining $\alpha_d$ is that it 
is necessary to solve the unperturbed and perturbed equations for both the bra and ket states. This amounts to a substantial increase in the computational
efforts to perform calculations using the NCC method in comparison with the CC method. Complexities grow further to implement it in the spherical 
coordinate system along with the angular spherical tensor products. Due to recent demands to perform high accuracy calculations in the atomic 
systems, we have developed the NCC method in the four-component relativistic theory (RNCC method) adopting the spherical coordinate system and it has 
been applied for the first time to calculate EDM and $\alpha_d$ values of the $^{199}$Hg atom \cite{sahoo4}. In this work, we apply the RNCC method to 
find out $\alpha_d$ of the Cd atom and compare the result with the other theoretical and experimental values. Furthermore, we also estimate this quantity
in the finite-field approach using the four-component Dirac-Coulomb (DC) Hamiltonian in the multi-reference coupled-cluster (MRCC) program \cite{MRCC}. 
By assessing various uncertainties and checking consistencies in the results from different methods at various levels of approximations, a precise value 
of $\alpha_d$ has been given. We also elucidate trends of correlation effects in the determination of this quantity by comparing intermediate results 
from a number of lower-order many-body methods and from different RCC and RNCC terms. In fact, there exists another novel CC approach for the determination 
of polarizabilities by evaluating the second derivative of energies \cite{Rozyczko}. However, development of such method using spherical coordinate system
is not straightforward and it will require one-more order expansion of (R)CC operators. This will give three different perturbed (R)CC operators similar 
to the approach described in Ref. \cite{shukla} for studying EDMs and it will lead to handling complicated tensor products to
account for the angular momentum couplings in the calculations of the perturbed wave functions.

 The remaining part of the paper is organized as follows: In the next section, we give briefly the theory of the atomic dipole polarizability. In 
Sec. \ref{sec-meth}, we describe the RCC and RNCC theories and then, discuss and present the results in Sec. \ref{sec-res}. We mention our 
conclusions in Sec. \ref{sec-sum}. Unless stated otherwise, we use  atomic units (a.u.) throughout the paper.

\section{Theory}\label{sec-theo}

The energy of the ground state of an atom in the presence of an external weak electric field of strength $\vec{\mathcal E}$ can be expressed in the perturbation theory as  \cite{manakov,bonin}
\begin{eqnarray}\label{polar}
E_0(|\vec{\mathcal E}|) = E_0(0) - \frac{\alpha_d}{2} |\vec{\mathcal E}|^2 - \dots,
\label{eqalp}
\end{eqnarray}
where $E_0(0)$ is the energy of the state in the absence of the electric field and $\alpha_d$ is known as the dipole polarizability of the state. It is obvious from the above expression that $\alpha_d$ 
can be determined by evaluating the second-order differentiation of $E_0(|\vec{\mathcal E}|)$ with a small magnitude of electric field $\vec{\mathcal E}$ as
\begin{equation}
 \alpha= - \left ( \frac{ \partial^2 E_0 (|\vec{\mathcal E}|)}{\partial |\vec{\mathcal E}| \partial |\vec{\mathcal E}|} \right )_{|\vec{\mathcal E}|=0} .
\end{equation}
This procedure is known as finite-field approach for evaluating $\alpha_d$ which involves calculations of $E_0(|\vec{\mathcal E}|)$ after including the interaction Hamiltonian $H_{int} = - \vec{\mathcal E} \cdot
\vec{D}$ with the atomic Hamiltonian. For achieving numerical stability in the result, it would be necessary to repeat the calculations by considering a number of $|\vec{\mathcal E}|$ values.

To estimate $\alpha_d$ in the spherical coordinate system, we can expand the ground state wave function of the atom in the presence of weak electric field as
\begin{eqnarray}
 |\Psi_0 \rangle = |\Psi_0^{(0)} \rangle + |\vec{\mathcal E}| |\Psi_0^{(1)} \rangle + \cdots
\end{eqnarray}
with $|\Psi_0^{(0)} \rangle$, $|\Psi_0^{(1)} \rangle$ etc. are the ground state wave function in the absence of the electric field, its first-order correction in the presence of electric field,
and so on. From the second-order perturbation expansion, we get
\begin{eqnarray}
\alpha_d &=& \frac{2}{\langle \Psi_0^{(0)}| \Psi_0^{(0)} \rangle } \sum_{I\ne 0}\frac{\langle \Psi_0^{(0)}|D|\Psi_I^{(0)} \rangle \langle \Psi_I^{(0)}|D|\Psi_0^{(0)} \rangle }{E_0^{(0)}(0)-E_I^{(0)}(0)} \nonumber \\
  &=& \frac{2}{\langle \Psi_0^{(0)}| \Psi_0^{(0)} \rangle } \sum_{I\ne 0}\frac{|\langle \Psi_0^{(0)}|D|\Psi_I^{(0)} \rangle|^2}{E_0^{(0)}(0)-E_I^{(0)}(0)},
\label{eq2}
\end{eqnarray}
where $|\Psi_I^{(0)} \rangle$ are the excited states of the atom  with energies $E_{I}^{(0)}(0)$. Allowing a mathematical formulation, we can express the first-order perturbed wave function of 
$|\Psi_0^{(0)} \rangle$ due to $D$ as
\begin{eqnarray}
 |\Psi_0^{(1)} \rangle = \sum_{I\ne 0} |\Psi_I^{(0)} \rangle \frac{\langle \Psi_I^{(0)}|D|\Psi_0^{(0)} \rangle }{E_0^{(0)}(0)-E_I^{(0)}(0)} .
\label{fowv}
\end{eqnarray}
Thus, the expression for $\alpha_d$ can be written as \cite{sahoo3}
\begin{equation}
\alpha_d = 2 \frac{\langle \Psi_0^{(0)}|D|\Psi_0^{(1)} \rangle}{ \langle \Psi_0^{(0)}| \Psi_0^{(0)} \rangle }.
\label{eqspd}
\end{equation}
In the {\it ab initio} approach, the above first-order perturbed wave function $|\Psi_0^{(1)} \rangle$ can be obtained as the solution to the
following inhomogeneous equation \cite{sahoo3}
\begin{eqnarray}
(H-E_0^{(0)}) |\Psi_0^{(1)} \rangle &=& -D|\Psi_0^{(0)} \rangle .
\label{eqfs}
\end{eqnarray}
This is equivalent to Dalgarno's approach \cite{dalgarno1,dalgarno2} except the fact that the solution for the above first-order perturbed equation has to be obtained for the dipole operator
$D$ in place of the interaction Hamiltonian $H_{int}$. Though dimension of $\vec{D}$ and $H_{int}$ are not same, but mathematically the solution of $|\Psi_0^{(1)} \rangle$ in Eq. (\ref{eqfs})
can give rise to the expression for $\alpha_d$ that is equivalent to Eq. (\ref{eq2}). Further, we can express
\begin{equation}
\alpha_d = \frac{1}{|\vec{\mathcal E}|} \frac{\langle \Psi_0|D|\Psi_0 \rangle}{ \langle \Psi_0| \Psi_0 \rangle },
\label{eqext}
\end{equation}
when $|\Psi_0 \rangle$ is evaluated only up to linear in $|\vec{\mathcal E}|$ correction.

\section{Methods for calculations}\label{sec-meth}

The exact wave function in the (R)CC theory is expressed as \cite{cizek}
\begin{equation} \label{exp}
| \Psi_0 \rangle = e^{\hat{T}} | \Phi_0^N \rangle
\end{equation}
where $|\Phi_0^N \rangle$ is the reference determinant, obtained using the $V^N$ potential of the $[4d^{10}5s^2]$ configuration of Cd in the Dirac-Hartree-Fock (DHF) method and $\hat T$ is known
as the (R)CC excitation operator given by
\begin{eqnarray} \label{Texp}
\hat T  &=&  \sum\limits_{k=1}^N  \hat T_k = \sum\limits_{\stackrel{a_1 < a_2 \dots < a_k}{i_1 < i_2 \dots < i_k}} t^{a_1 a_2 \dots a_k}_{i_1 i_2 \dots i_k}   a^+_1 i^-_1 a^+_2 i^-_2 \dots a^+_k i^-_k , \ \ \ \
\end{eqnarray}
where $+$ and $-$ superscripts on the second quantization operators represent for the creation and annihilation of electrons in the virtual (denoted by $a$) and occupied (denoted by $i$) orbitals,
respectively, and $t$ are the amplitudes in the excitation process in an $N$ electron system. The (R)CC approaches considering up to $T_N$ operators with $N = 2, 3, 4, \dots$, known as the (R)CC 
singles and doubles (CCSD), (R)CC singles, doubles, and triples (CCSDT), (R)CC singles, doubles, triples, and quadruples (CCSDTQ), etc. methods constitute a hierarchy, which converges to the exact 
solution of the wave function in the given one-particle basis set.

 The amplitudes $t$ of the (R)CC operators are obtained by projecting bra determinants $\langle \Phi^{a_1 a_2 \dots a_k}_{i_1 i_2 \dots i_k} | e^{-\hat{T}}=$ $\langle \Phi_0^N | a^+_1 i^-_1 a^+_2 i^-_2 \dots  a^+_k i^-_k e^{-\hat{T}} $ from the
left of the Schr\"odinger equation $\hat H |\Psi_0 \rangle = E_0 |\Psi_0 \rangle$, with the ground state energy $E_0$, as \cite{bartlett,crawford}
\begin{eqnarray}\label{CCeq}
\langle \Phi^{a_1 a_2 \dots a_k}_{i_1 i_2 \dots i_k} \vert \overline{H} \vert \Phi_0^N \rangle &=& E_0 \delta_{k,0}, \;\;\;\; (k=1, \dots N) ,
\end{eqnarray}
where $ \overline{H}=e^{-\hat{T}} \hat H e^{\hat{T}}=(\hat H e^{\hat{T}})_c$ for the subscript $c$ means connected terms with the atomic Hamiltonian $\hat H$.

We also perform calculations starting with the $V^{N-2}$ potential for the $[4d^{10}]$ configuration of Cd in the DHF wave function calculation by expressing
\begin{eqnarray}
| \Psi_0 \rangle = \hat{W} e^{\hat{T}} | \Phi_0^{N-2} \rangle ,
\end{eqnarray}
with $\hat T  =  \sum\limits_{k=1}^{N-2} \hat T_k$ and the doubly valence electron attachment operator $\hat{W}=\sum_{k=1}^{N-2} \hat{W}_k$ is defined as
\begin{eqnarray}
\hat{W} &=&  \sum \limits_{\stackrel{a_3 < a_4 \dots < a_k}{i_3 < i_4 \dots < i_k}} w^{a_3 a_4 \dots a_k}_{i_3 i_4 \dots i_k} a_1^+ a_2^+ a^+_3 i^-_3  \dots a^+_k i^-_k ,
\end{eqnarray}
for the corresponding amplitude $w$. In this approach, we evaluate the double attachment energy $\Delta E_{att}^2$ in the equation-of-motion framework as
\begin{eqnarray}
 \left [\overline{H},\hat{W} \right ] | \Phi_0^{N-2} \rangle = \Delta E_{att}^2 \hat{W} | \Phi_0^{N-2} \rangle .
\end{eqnarray}

In the finite-field procedure, we first calculate the total energy by considering the DC Hamiltonian, $H \equiv H^{DC}$, of the atom given by
\begin{eqnarray}
H^{DC} &=&\sum_i  \left [ c\mbox{\boldmath$\alpha$}_i \cdot \textbf{p}_i+ \beta_i c^2+ V_{nuc} (r_i) + \sum_{j \ge i} \frac{1}{r_{ij}} \right ],  \ \ \ \
\end{eqnarray}
where $\mbox{\boldmath$\alpha$}$ and $\beta$ are the Dirac matrices, $c$ is the speed of light, and $V_{nuc}(r)$ is the nuclear potential energy in the atom. We use the MRCC program \cite{MRCC} 
to perform the RCC calculations in the finite-field approach. The one-body and two-body integrals were generated using the DIRAC package \cite{Dirac} for the MRCC program. We evaluate energies 
$E_0(|\vec{\mathcal E}|)$ by considering the total Hamiltonian as $H \equiv H^{DC}+H_{int}$ using a number of $|\vec{\mathcal E}|$ values as 0.0, 0.0005, 0.001, and 0.002 in a.u. to 
estimate $\alpha_d$.

In the finite-field approach it is not required to define separate $\hat T$ operators of the RCC method in the absence and presence of the interaction Hamiltonian 
$H_{int}$ in the atomic Hamiltonian. However, it is necessary to do so in the perturbative approach of the RCC method. For this purpose, we express the RCC wave 
function in this case as
\begin{eqnarray}
 |\Psi_0 \rangle =e^{\hat{T}^{(0)}+ |\vec{\mathcal E}| \hat{T}^{(1)}} |\Phi_0^N \rangle,
\end{eqnarray}
where $\hat{T}^{(0)}$ represents for the RCC operator that accounts for electron correlation effects due to the electromagnetic interactions only and $\hat{T}^{(1)}$ takes care of correlation effects due to
both the electromagnetic interactions and the $D$ operator, respectively, to all-orders. In the perturbative expansion, this corresponds to
\begin{eqnarray}
 |\Psi_0^{(0)} \rangle = e^{\hat{T}^{(0)}} |\Phi_0^N \rangle
\ \ \ \ 
\text{and}
\ \ \ \
 |\Psi_0^{(1)} \rangle = e^{\hat{T}^{(0)}} \hat{T}^{(1)} |\Phi_0^N \rangle.
 \label{eqt1}
\end{eqnarray}
Both $|\Psi_0^{(0)} \rangle$ and $|\Psi_0^{(1)} \rangle$ can be determined by obtaining amplitudes of the $\hat{T}^{(0)}$ and $\hat{T}^{(1)}$ RCC operators. The amplitude determining equation
for $\hat{T}^{(0)}$ is same as Eq. (\ref{CCeq}) for the DC Hamiltonian. The $\hat{T}^{(1)}$ amplitude determining equation is given by \cite{yashpal1,yashpal2,sahoo1,sahoo2}
\begin{eqnarray}
 \langle \Phi^{a_1 a_2 \dots a_k}_{i_1 i_2 \dots i_k} \vert \overline{H}^{DC} \hat{T}^{(1)}+ \overline{D} \vert \Phi_0^N \rangle =0 \label{eq37} .
\end{eqnarray}
It to be noted that for solving the amplitudes of $\hat{T}^{(0)}$, the projected $\langle \Phi^{a_1 a_2 \dots a_k}_{i_1 i_2 \dots i_k} \vert$ determinants have to be even parity whereas they are the
odd-parity for the evaluating the $\hat{T}^{(1)}$ amplitudes. In the CCSD method approximation, we denote the RCC operators as
\begin{eqnarray}
 {\hat T}^{(0)}= T_1^{(0)} + T_2^{(0)} \ \ \ \ \text{and} \ \ \ \  {\hat T}^{(1)}= T_1^{(1)} + T_2^{(1)},
\end{eqnarray}
where subscripts $1$ and $2$ stands for the singles and doubles excitations, respectively.

After obtaining these solutions, we can evaluate $\alpha_d$, following Eq. (\ref{eqspd}), as \cite{yashpal2,sahoo4}
\begin{eqnarray}
\alpha_d &=& \frac{1}{|\vec{\mathcal E}|} \frac{\langle\Phi_0^N | e^{T^{\dagger}} D e^T | \Phi_0^N \rangle }
                  {\langle\Phi_0^N | e^{T^{\dagger}} e^T | \Phi_0^N \rangle } = \frac{1}{|\vec{\mathcal E}|} \langle\Phi_0^N | e^{T^{\dagger}} D e^T | \Phi_0^N \rangle_{fc} \nonumber \\
                  &=& 2 \langle\Phi_0^N | e^{T^{(0)\dagger}} D e^{T^{(0)}} T^{(1)} | \Phi_0^N \rangle_{fc} ,
\label{eqccx}
\end{eqnarray}
where $fc$ stands for the fully-contracted terms. The above expression contains a non-terminating series $e^{T^{\dagger (0)}} D e^{T^{(0)}}$. This is computed self-consistently as discussed in 
Refs. \cite{yashpal2,sahoo4}.

It is worth mentioning two things here. First, the normalization factor in Eq. (\ref{eqccx}) appears 
explicitly in the PRCC method while, as shown above, it CANCELS out in our approach. Secondly, partial triple excitation are included in the PRCC(T) method by defining a perturbative operator as
\begin{eqnarray}
 T_3^{(1),pert}= \frac{1}{3!}\sum_{abc,pqr}  \frac{ ( H^{DC} T_2^{(1)})_{abc}^{pqr} }{\epsilon_a + \epsilon_b+\epsilon_c-\epsilon_p -\epsilon_q -\epsilon_r} 
 \label{eqt31}
\end{eqnarray}
with $a,b,c$ and $p,q,r$ subscripts denoting for the occupied and unoccupied orbitals, respectively, and considering it as a part of $T^{(1)}$ in their property evaluating expression like 
Eq. (\ref{eqccx}). To make a similar analysis, we also include the above operator in Eq. (\ref{eqccx}) in our method to estimate the partial triples effects to the CCSD method and refer this 
approach as the CCSD(T) method in order to be consistent with the notation of Ref. \cite{sidh}. However, it should be noted that $T_1^{(1)}$ operator is the dominant over $T_2^{(1)}$ in 
the perturbative approach owing to one-body form of the $D$ operator. Thus, the above approach cannot estimate triples effects rigorously. On the other hand, $T_2^{(0)}$ DOMINATES over the $T_1^{(0)}$ operator due to THE two-body nature of 
the Coulomb interaction. Therefore, it is necessary to include important triples effects through the $T^{(0)}$ operator. We define another triple excitation operator as
\begin{eqnarray}
 T_3^{(0),pert}= \frac{1}{3!}\sum_{abc,pqr}  \frac{ ( H^{DC} T_2^{(0)})_{abc}^{pqr} }{\epsilon_a + \epsilon_b+\epsilon_c-\epsilon_p -\epsilon_q -\epsilon_r} 
 \label{eqt30}
\end{eqnarray}
and consider it as a part of the $T^{(0)}$ operator. Moreover, we include both the $T_3^{(0),pert}$ and $T_3^{(1),pert}$ operators in the amplitude determining equations as well as 
in the property evaluating expression given by Eq. (\ref{eqccx}). We refer to this procedure as the CCSDTp method in the present work.

For the calculation of $\alpha_d$ using Eq. ($\ref{eqccx}$) in the (R)CC method, the bra state was used as the complex conjugate of the ket state. In the (R)NCC method, however, the ket state is
determined in the same way as the (R)CC method but another bra state is used for the corresponding ket state $| \Psi_0 \rangle$ and is expressed by \cite{arponen,bishop}
\begin{eqnarray}
 \langle \tilde{\Psi}_0 | = \langle \Phi_0^N | (1+\hat{\Lambda}) e^{-\hat{T}} ,
\end{eqnarray}
where $\hat \Lambda$ is a de-excitation operator defined as
\begin{eqnarray} \label{Texp}
\hat \Lambda  &=&  \sum\limits_{k=1}^N  \hat \Lambda_k = \sum\limits_{\stackrel{i_1 < i_2 \dots < i_k}{a_1 < a_2 \dots < a_k}} {\tilde t}^{i_1 i_2 \dots i_k}_{a_1 a_2 \dots a_k}  i^+_1 a^-_1 i^+_2 a^-_2 \dots i^+_k a^-_k , \ \ \ \
\end{eqnarray}
where ${\tilde t}$ represents amplitude for the corresponding de-excitation operator. The following bi-orthogonal condition between these two states is evident
\begin{eqnarray}
 \langle \tilde{\Psi}_0 | \Psi_0 \rangle = \langle \Phi_0^N | (1+\hat \Lambda) e^{-\hat T} e^{\hat T} |\Phi_0^N \rangle =1 .
\end{eqnarray}
If $\langle \tilde{\Psi}_0 |$ has the same eigenvalue $E_0$ of $| \Psi_0 \rangle$, then $\langle \tilde{\Psi}_0 |$ can be used in place of $\langle \Psi_0 |$ in the calculation of an expectation value.
This choice of bra in the (R)NCC method also satisfies the Hellmann-Feynman equation \cite{bishop} in contrast to the ordinary (R)CC method. This is attained with the following prerequisite condition
\begin{eqnarray}
\langle \Phi_0^N | \hat \Lambda \overline{H}|\Phi_0^N \rangle= 0 .
\end{eqnarray}
Indeed, this is the case as per the amplitude solving equation Eq. (\ref{CCeq}) of $\hat T$. Now it is necessary to expand the $\hat \Lambda$ operator perturbatively like the $\hat T$ operator to obtain 
the first-order perturbed wave function of the bra state for the 
evaluation of $\alpha_d$. Thus, we write
\begin{eqnarray}
 \langle \tilde{\Psi}_0 | &=& \langle \tilde{\Psi}_0^{(0)} | +  |\vec{\mathcal E}| \langle \tilde{\Psi}_0^{(1)} | + \cdots \nonumber \\
 &=& \langle \Phi_0^N | (1+\Lambda^{(0)} + \lambda \Lambda^{(1)}+\cdots) e^{-(T^{{0}} + |\vec{\mathcal E}| T^{(1)})} . \ \ \ \
\end{eqnarray}
Equating to terms of zeroth and linear in $|\vec{\mathcal E}|$, we get
\begin{eqnarray}
 \langle \tilde{\Psi}_0^{(0)} | = \langle \Phi_0^N | (1+\Lambda^{(0)}) e^{-T^{(0)}}
\end{eqnarray}
and
\begin{eqnarray}
\langle \tilde{\Psi}_0^{(1)} | = \langle \Phi_0^N | \left [ (1+\Lambda^{(0)}) T^{(1)} + \Lambda^{(1)}) e^{-T^{(0)}} \right ] , \ \ \ \ \
\end{eqnarray}
respectively. In order to determine these wave functions, amplitudes of the $\Lambda^{(0)}$ and $\Lambda^{(1)}$ RNCC operators are obtained by solving the following equations \cite{sahoo4}
\begin{eqnarray}
\langle \Phi_0^N \vert \Lambda^{(0)} \overline{H}^{DC} + \overline{H}^{DC}  \vert \Phi^{a_1 a_2 \dots a_k}_{i_1 i_2 \dots i_k} \rangle =0
\end{eqnarray}
and
\begin{eqnarray}
\langle \Phi_0^N \vert \left [ \Lambda^{(1)} \overline{H}^{DC}  + (1+ \Lambda^{(0)}) \left \{ \overline{D} + (\overline{H}^{DC} T^{(1)})_{c} \right \} \right ], \nonumber \\ \vert \Phi^{a_1 a_2 \dots a_k}_{i_1 i_2 \dots i_k} \rangle =0
\end{eqnarray}
respectively. It can be noticed that the above equations contain more terms than the $T^{(0/1)}$ amplitude solving equations. Since it contains more non-linear terms, it means efforts to code the (R)NCC 
method are more than twice compared to the (R)CC method.

Knowing amplitudes of the RCC and RNCC operators, we can evaluate $\alpha_d$ using the expression as \cite{sahoo4}
\begin{eqnarray}
\alpha_d &=&  \frac{1}{|\vec{\mathcal E}|}\frac{\langle \Psi_0 | D | \Psi_0 \rangle}{\langle \Psi_0 | \Psi_0 \rangle} = \frac{1}{|\vec{\mathcal E}|} \frac{\langle \tilde{\Psi}_0 | D | \Psi_0 \rangle}{\langle \tilde{\Psi}_0 | \Psi_0 \rangle} \nonumber \\
  &=&  \langle\Phi_0^N | (1+\Lambda) e^{-T} D e^T | \Phi_0^N \rangle_{fc}  \nonumber \\
                  &=& \lambda \langle\Phi_0^N | (1+\Lambda^{(0)})  \overline{D} T^{(1)} + \Lambda^{(1)} \overline{D} | \Phi_0^N \rangle_{fc} .
\label{eqnccx}
\end{eqnarray}
This expression does not have any non-terminating series in contrast to the expression given by Eq. (\ref{eqccx}) and the normalization of the wave function does not appear in a natural way. Since $D$ is an one-body operator, the above 
expression will also have a fewer terms for the evaluation of $\alpha_d$ as the compensation to the extra calculations for the amplitudes of the $\hat \Lambda$ operator. Nevertheless, it is desirable to 
obtain consistent values for $\alpha_d$ in the approximated RCC and RNCC methods in order to justify reliability in the theoretical calculation of the $\alpha_d$ value. We define
the NCC method with the singles and doubles excitations approximation as the NCCSD method and the NCC method with the singles, doubles and important perturbative 
triples excitations approximation as the NCCSD(T) method in this work. 

We also perform calculations employing many-body perturbation theory considering $n$ orders, say, of residual Coulomb interactions (designated as MBPT(n) method) to fathom the propagation of electron correlation effects
from lower- to all-order many-body methods. In the finite-field approach, the commonly known MBPT(n) approach has been adopted while we define the unperturbed and the first-order perturbed wave operators 
in the wave function expansion approach as \cite{yashpal1}
\begin{eqnarray}
 |\Psi_0^{(n,0)} \rangle = \sum_{\beta=1}^{n} \Omega^{(\beta,0)} |\Phi_0^N \rangle
 \end{eqnarray}
and
\begin{eqnarray}
 |\Psi_0^{(n,1)} \rangle = \sum_{\beta=1}^{n-1} \Omega^{(\beta,1)} |\Phi_0^N \rangle,
\end{eqnarray}
respectively, where the first superscript index $n$ represents for order of residual Coulomb interactions and the second superscript 0/1 indicates presence of number of $D$ operator in the evaluation of 
these wave functions. In this framework, we evaluate $\alpha_d$ by \cite{yashpal1}
\begin{eqnarray}
\alpha_d &=& 2 \frac{\sum_{\beta=0}^{n-1} \langle \Phi_0^N| {\Omega^{(n-\beta,0)}}^{\dagger} D \Omega^{(\beta,1)} |\Phi_0^N \rangle}
{ \sum_{\beta=0}^{n-1} \langle \Phi_0^N| {\Omega^{(n-\beta,0)}}^{\dagger} \Omega^{(\beta,0)} |\Phi_0^N \rangle} .
\end{eqnarray}
It is worth noting that the MBPT(n) method in the perturbative formulation is equivalent to the MBPT(n-1) method of the finite-field approach as both involve up to the same orders of residual Coulomb interactions.

Also by perturbing the DHF orbitals to first-order by the $D$ operator and adopting a self-consistent procedure, we can include the core-polarization effects to all-orders in the  
RPA for the evaluation of $\alpha_d$ \cite{yashpal2}. In this approach, we express
\begin{eqnarray}
 \alpha_d  & = & 2 \langle \Phi_0^N| D \Omega_{RPA}^{(1)} |\Phi_0^N \rangle ,
\end{eqnarray}
where the perturbed $\Omega_{RPA}^{(1)}$ wave operator is defined in our earlier work \cite{yashpal2}. From the differences between the results obtained by the RPA and CCSD methods in the perturbative 
approach, we can find out contributions from the non-core-polarization correlations to all-orders.

We have estimated Breit interaction contribution by adding the following term \cite{breit} in the atomic Hamiltonian
\begin{eqnarray}
V_B(r_{ij})=-\frac{1}{2r_{ij}}\{\mbox{\boldmath$\alpha$}_i\cdot \mbox{\boldmath$\alpha$}_j+
(\mbox{\boldmath$\alpha$}_i\cdot\bf{\hat{r}_{ij}})(\mbox{\boldmath$\alpha$}_j\cdot\bf{\hat{r}_{ij}}) \} .
\end{eqnarray}
We also estimate contributions from the lower order vacuum polarization (VP) effects using the Uehling ($V_{U}(r)$) and Wichmann-Kroll ($V_{WK}(r)$) potential
energies and self-energy (SE) effects by including the corresponding potential energies due to the electric and magnetic form-factors that have been described 
in our earlier work \cite{sahoo5}. 

We use Gaussian type orbitals (GTOs) to construct the electron orbitals in the DHF method. The $k$th GTO in the basis expansion is defined as \cite{mohanty}
\begin{eqnarray}
 \chi_k(r)  &=& r^l e^{-\zeta_k r^2},
\label{anbas}
\end{eqnarray}
with the orbital quantum number $l$ and for an arbitrary parameter $\zeta_k$. Similarly, we use the Dyall's uncontracted correlated consistent double-, triple-, quadruple-$\zeta$ GTO basis sets \cite{Basis-Cd},
which are referred to as $X\zeta$, where $X$=2, 3, and 4, respectively, in the DIRAC package \cite{Dirac} to generate the one-body and two-body integrals for the MRCC program \cite{MRCC}. Each shell is augmented by two additional diffuse functions (d-aug) and the exponential coefficient of the augmented function is
calculated based on the following formula
\begin{equation}
\zeta_{N+1}= \left [\frac{\zeta_N}{\zeta_{N-1}} \right ]\zeta_{N} ,
\end{equation}
where $\zeta_{N}$ and $\zeta_{N-1}$ are the two most diffuse exponents for the respective atomic-shells in the original GTOs. For the spherical coordinate system in the perturbative approach of $\alpha_d$
calculation, we construct $\zeta_k$ using the even tempering condition defining as
\begin{eqnarray}
\zeta_k &=& \zeta_0 \eta^{k-1},
\label{evtm}
\end{eqnarray}
with two unknown parameters $\zeta_0$ and $\eta$. We have chosen $\zeta_0$ parameter as 0.00715, 0.0057, 0.0072, 0.0052, 0.0072, and 0.0072 while $\eta$ 
parameter as 1.92, 2.04, 1.97, 2.07, 2.54 and 2.54 for orbitals with $l=$0, 1, 2, 3, 4 and 5, respectively, after optimizing the single particle orbital
energies.

\begin{table}[t]
\caption{A summary of $\alpha_d$ values in $ea_0^3$ of the Cd atom from various calculations and measurement is presented. We give results from the finite-field approach and perturbing
wave function approach in separate columns. As can be seen trends are different in both the approaches. Calculations carried out using (R)CC variant methods are supposed to be more reliable. The CCSD and PRCC
methods (and their variants) are equivalent, but differ only in the implementation technique. Uncertainties are quoted within the parentheses and references from other works are cited beside the 
corresponding results. The recommended value from the present work is quoted at the bottom of the table.}
\begin{ruledtabular}
\begin{tabular}{lcc}
  &       &  \\
       & Finite-field  & Perturbation  \\
\hline
& & \\
   \multicolumn{3}{c}{$\alpha_d$ values from this work}   \\
DHF     &  63.657  & 49.612   \\
MBPT(2) &  37.288  & 50.746   \\
MBPT(3) &          & 37.345                       \\
RPA     &          & 63.685                       \\
CCSD$^{\star}$ & 47.618 &    \\
CCSD    &  48.073  & 45.494      \\
NCCSD   &          & 44.804     \\
CCSD(T) &    &   45.586   \\
CCSDTp  &    &   46.289   \\
NCCSD(T)&          & 45.603      \\
CCSDT   &  45.852  & \\
CCSDTQ  &  45.927  &  \\
  & & \\
$\Delta$Breit      & & 0.105    \\
$\Delta$QED   & & 0.105  \\
Final   & 46.015(203)  &  46.0(5)  \\
\hline
&   &    \\
 \multicolumn{3}{c}{$\alpha_d$ values from previous calculations}   \\
 DHF    &  62.78 \cite{kello}, 63.37 \cite{seth}    &  49.647 \cite{yashpal2} \\
 MBPT(2) & 39.14 \cite{kello}, 38.52 \cite{seth}    &  \\
 MBPT(3) &  45.97 \cite{kello}, 45.86 \cite{seth}                   & 35.728 \cite{yashpal2} \\
 MBPT(4) &  45.06 \cite{kello}, 47.10 \cite{seth}   & \\
 CICP    &   & 44.63 \cite{ye}    \\
 CCSD    & 48.43 \cite{kello}, 48.09 \cite{seth} &  45.898 \cite{yashpal2} \\
 CCSD(T) & 46.80 \cite{kello}, 46.25 \cite{seth}  & \\
 PRCC    &         &  49.15 \cite{sidh} \\
PRCC(T) &          &  49.24 \cite{sidh} \\
\hline
&   &    \\
Experiment  & \multicolumn{2}{c}{$49.65 \pm 1.49 \pm 0.16$ \cite{goebel}} \\
            & \multicolumn{2}{c}{45.3 \cite{goebel1,hohm1}} \\
Recommended & \multicolumn{2}{c}{ 46.02(50) } \\
\end{tabular}
\end{ruledtabular}
\label{tab1}
\end{table}

\begin{figure}[t]
\begin{center}
\includegraphics[width=9.5cm,height=9.0cm]{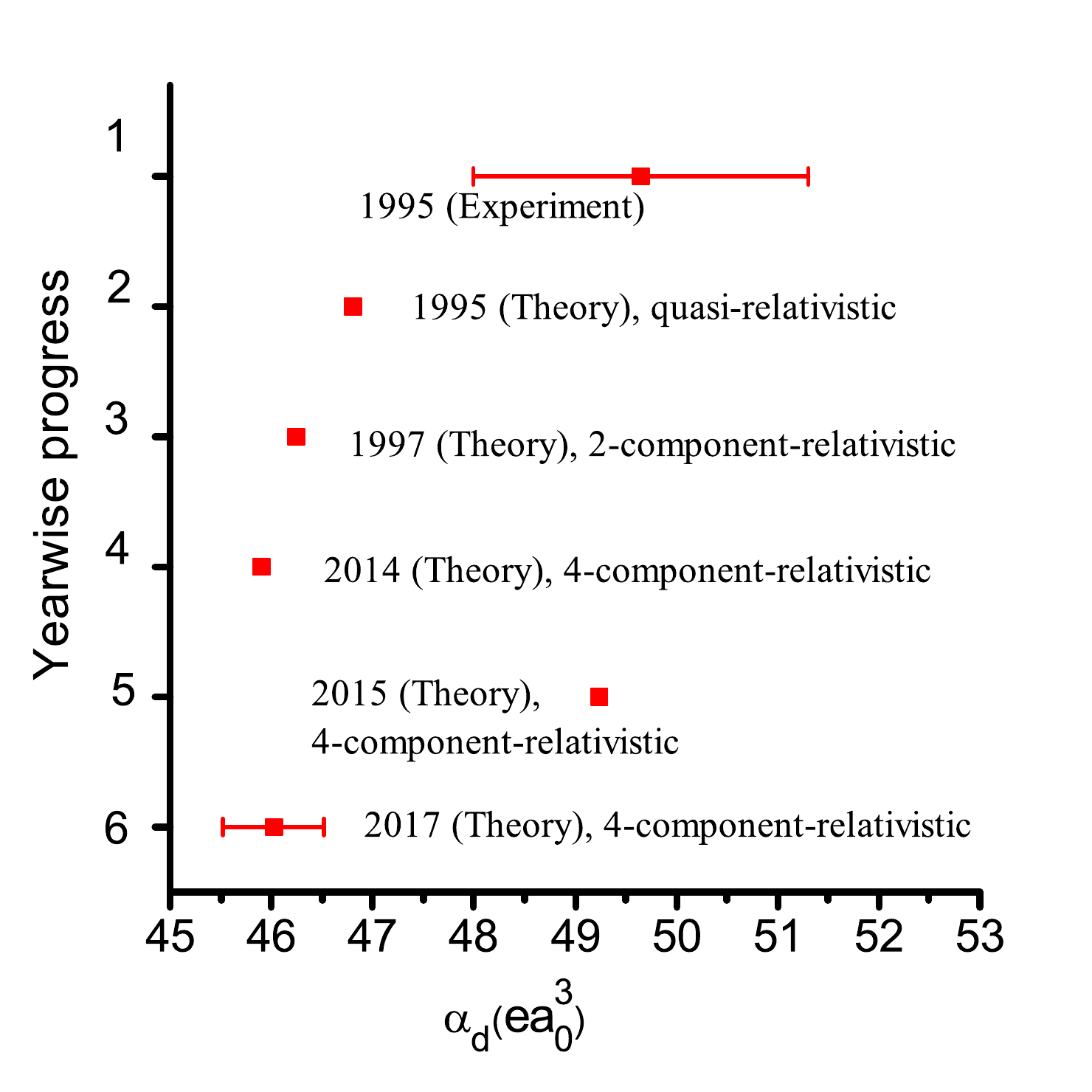}
\end{center}
\caption{(Color online) Year-wise progress of the $\alpha_d$ value (in $ea_0^3$) of the Cd atom from various works.}
\label{fig1}
\end{figure}

\section{Results and Discussion}\label{sec-res}

In Table \ref{tab1}, we list the $\alpha_d$ values of the Cd atom obtained using various many-body methods and also from the measurements. Though we quote in this table two experimental values \cite{goebel,goebel1},
but they are obtained from the same experimental set up. The most precise measurement is reported as $49.65(1.65) \ e a_0^3$ \cite{goebel}, while we have been informed \cite{hohm1} that a value of 45.3 
$ea_0^3$ for the static polarizability can be inferred from the preliminary experimental data of dynamic polarizabilities reported using the dispersive Fourier transform spectroscopy analysis \cite{goebel1}. 
Following these measurements, $\alpha_d$ value of Cd was theoretically studied by Kell\"o and Sadlej using the nonrelativistic CC theory and the first-order basis sets in the finite-field approach. They had obtained the results as 57.39 $ea_0^3$ and 55.36 $ea_0^3$ in the CCSD and CCSD(T) approximations, respectively. After inclusion of quasi-relativistic correction through the 
mass-velocity and Darwin terms, the final CCSD(T) value was quoted as 46.80 $ea_0^3$. In fact, this study had suggested for the first time about large contributions from the triples and relativistic effects to 
$\alpha_d$ of Cd. This result was slightly smaller than the above precise measurement. Later, this trend was confirmed by Seth {\it et al}. \cite{seth} employing the CCSD(T) method. But they had used pseudo-potential 
in the two-component relativistic Hamiltonian in their calculations. After few years of this work, a four-component relativistic theory with the semi-empirical core-potential in the configuration interaction (CICP) approach was employed and reported a value
of 44.63 $ea_0^3$ \cite{ye}. Apart from this, it uses a sum-over-states approach mentioned by Eq. (\ref{eq2}) with the $V^{N-2}$ potential. In the same work, the authors also give calculated values of $\alpha_d$ for
the Zn and Hg atoms using the CICP method and the results were found to be quite off from their respective experimental values. In the year 2014, we had employed our perturbative RCC theory in the CCSD method 
approximation to estimate its value using the four-component relativistic DC Hamiltonian and accounting for correction from the Breit interaction \cite{yashpal2}. The obtained result 45.86(15) $ea_0^3$  
was close to the previous CCSD(T) calculations in the finite-field approach \cite{kello,seth}. Following our work, Chattopadhyay {\it et al}. had applied their PRCC(T) method in the four-component 
relativistic theory and reported the $\alpha_d$ value as 49.24 $ea_0^3$ \cite{sidh}. This theoretical result was very close to the experimental value of $49.65(1.65) \ e a_0^3$. The difference between both the calculations was 
attributed to the inexactness in the evaluation of the RCC expression of Eq. (\ref{eqccx}) in these works. In fact, about 15\% contribution of total value is added due to the normalization of the wave function in 
Ref. \cite{sidh}, while we had omitted this contribution arguing its cancellation with the disconnected part of the numerator \cite{yashpal2}. In this work, we find values from both the CCSD and NCCSD methods in the perturbative approach are very close to each other. In 
fact, the results are becoming even closer in the CCSD(T) and NCCSD(T) methods. This certainly demonstrates normalization of the wave function does not contribute to the $\alpha_d$ value of the ground state 
of a closed-shell atomic system in the RCC theory framework. Moreover, our results from the finite-field approach using the CCSDT and CCSDTQ methods with the four-component relativistic DC Hamiltonian
are also close to the results of the perturbative CCSD(T) and NCCSD(T) methods. Even though both the procedures, finite-field and perturbative, adopted here are very different, but good agreement between the results 
obtained from these calculations strongly advocate for their reliability. We recommend its value as 46.02(50) $ea_0^3$ by taking into account various uncertainties as discussed below. We show gradual progress in the experimental and theoretical results 
over the years in Fig. \ref{fig1}, which clearly indicates most of the theoretical results agree with each other except the values from the PRCC and PRCC(T) methods.

\begin{table}[t]
\caption{Demonstration of convergence of result in the perturbative approach with different set of active orbitals 
in the CCSD method.}
\begin{ruledtabular}
\begin{tabular}{lcc}
 Basis set & Active orbitals & Result \\\hline
  & & \\
 Set I  &  1-15$s$, 2-13$p$, 3-13$d$, 4-10$f$ & 46.034\\
 Set II  &  1-15$s$, 2-15$p$, 3-15$d$, 4-15$f$ & 45.872 \\
 Set III  &  1-17$s$, 2-17$p$, 3-17$d$, 4-16$f$ & 45.758 \\
 Set IV  & 1-17$s$, 2-17$p$, 3-17$d$, 4-16$f$, 5-14$g$  & 45.494 \\ 
 Set V  & 1-21$s$, 2-21$p$, 3-21$d$, 4-18$f$, 5-16$g$  & 45.494 \\
 Set VI  & 1-21$s$, 2-21$p$, 3-21$d$, 4-18$f$, 5-16$g$, 6-10$h$  & 45.494 \\
 Set VII  & 1-21$s$, 2-21$p$, 3-21$d$, 4-18$f$, 5-16$g$, 6-12$h$  & 45.494 \\
 \end{tabular}
\end{ruledtabular}
\label{tab02}
\end{table}

In the above table, we also give corrections from the Breit (quoted as $\Delta$Breit) and QED (quoted as $\Delta$QED) interactions explicitly by estimating them from RPA.
We found these contributions are negligibly small. Therefore, 
uncertainties to $\alpha_d$ can come mainly from the finite-size basis used in the calculations and contributions from the neglected higher level excitations. The results obtained by us earlier in Ref. \cite{yashpal2}
and in this work by the CCSD method differ slightly due to use of different basis functions. We had also estimated contributions from the partial triples but only through the perturbed $T_3^{(1),pert}$ RCC operator
including in the amplitude determining equations of the CCSD method and were referred to as the CCSDpT method \cite{yashpal2}. In this work, we have estimated these contributions more rigorously after including 
triples effects through the unperturbed and perturbed RCC operators as well as estimating contributions from the $T_3^{(1),pert}$ RCC operator in Eq. (\ref{eqccx}). In Table \ref{tab02}, we demonstrate convergence 
of the result obtained using perturbative approach in the CCSD method. After accounting for uncertainties, we find that 
$\alpha_d=46.0(5) \ ea_0^3$ in the wave function perturbative approach. To assess uncertainties associated with our result obtained in the finite-field approach, we describe here how these calculations 
were performed systematically up to the CCSDTQ method. Contributions from different levels of excitations and inner core orbital correlations, that was neglected in the CCSDT and CCSDTQ methods, are listed in 
Table \ref{tab2}. Due to limited available computational resources, it was not possible to consider correlations among all the core electrons in the CCSDT and CCSDTQ methods using the MRCC program \cite{MRCC}. 
Thus, we perform first the CCSD calculations using the $4\xi$ basis but considering electrons only from the $3d$, $4s$, $4p$ and $4d$ shells (given as $\alpha_d^{\text{CCSD}}$). Contributions from the inner core 
orbitals were estimated using the $2\xi$ basis in the CCSD method and given as $\alpha_d^{\text{Core}}$. We had, then, performed calculations using the $4s$, $4p$ and $4d$ orbitals in the 
CCSD and CCSDT methods. The difference is quoted as triples contribution (given as $\alpha_d^{\text{T}}$) and uncertainty due to exclusion of other orbitals in the CCSDT method is estimated by scaling their 
contributions in the CCSD method. The quadruples effects are estimated using orbitals from the $4d$ shell alone again with the $2\xi$ basis (given as $\alpha_d^{\text{Q}}$) and the same value has been taken as 
the maximum possible uncertainty due to the quadruple excitations arising from the other less active inner orbitals. Details of these contributions along with their uncertainties can be found in Table \ref{tab2}. 
Adding all these uncertainties together, we anticipate $\alpha_d$ in the finite-field approach as 46.015(203) $ea_0^3$. This is in very good agreement with the value obtained in the perturbative wave function 
approach. Now taking into confidence on the estimated uncertainties from both the procedures, we have recommended optimistically the final $\alpha_d$ value of the Cd atom as 46.02(50) $ea_0^3$.

\begin{table}[t]
\caption{Breakdown of various contributions to $\alpha_d$ in $ea_0^3$ of Cd along with their uncertainties from the finite-field approach calculation in this work. Basis functions used in different steps 
are also mentioned for the clarity.}
\begin{ruledtabular}
\begin{tabular}{lcc}
 Source       & Contribution & Basis \\\hline
  & & \\
 $\alpha_d^{\text{CCSD}}$         & 47.678$\pm$0.096  & $4\xi$       \\
 $\Delta \alpha_d^{\text{T}}$     & $-1.370\pm$0.040  & $2\xi$        \\
 $\Delta \alpha_d^{\text{Q}}$     & $-0.075\pm$0.075  & $2\xi$        \\
 $\Delta \alpha_d^{\text{Core}}$  & $-0.176\pm$0.023  & $2\xi$        \\
 \end{tabular}
\end{ruledtabular}
\label{tab2}
\end{table}

It can also be noticed from Table \ref{tab1} that the trends of our finite-field results at the DHF value is very large and the MBPT(2) result is lower than the CCSD and CCSD(T) values. The reason for which the
DHF value is large in this case is understandable as it is obtained using the variational approach. Compared to the finite-field approach, the trends obtained at various levels of approximations in the perturbative
approach is completely different. In this formalism, the DHF method does not give the largest value since the procedure to estimate the expectation value in this case is not variational.  
RPA gives a very large value with respect to the DHF result implying core-polarization correlations are very strong in this system. The RPA value of the perturbative approach is close to the DHF value of the finite-field approach. The reason is DHF value in the finite-field approach 
includes orbital relaxation effect, which is explicitly taken care by RPA in the perturbative approach. As we had stated before, the MBPT(n) method approximations 
in the perturbative approach is equivalent to the MBPT(n-1) method approximation in the 
finite-field approach. This is why the MBPT(2) value of the finite-field approach matches with the MBPT(3) value of the perturbative approach. The above agreements between both the procedures support correct
implementation of the methods. Also, significant difference between the RPA and CCSD results suggest that there are also large contributions come from the all-order non-core-polarization effects. The final result is the outcome of the cancellation 
between these two contributions, and become closer to the DHF value of the perturbative approach. Another point to be realized that the inclusion of contributions from the triples excitations increase 
the value in the perturbative formalism in contrast to the finite-field approach.

\begin{table}[t]
\caption{Comparison of contributions to $\alpha_d$ in $ea_0^3$ among various RCC terms from our CCSD and NCCSD methods with the PRCC method of Ref. \cite{sidh}. Contributions from the h.c. terms 
are given separately in order to make a comparative analysis with the contributions from the bra terms of the NCCSD method. Contribution due to normalization factor of the wave function is given 
explicitly for the PRCC method. Contributions from the higher-order non-linear terms that are not mentioned here are given combining as ``Others''. As can be seen, contributions from various RCC terms in the CCSD and PRCC 
methods differ significantly.  Also, the bra terms of the NCCSD method give quite different 
values than the CCSD method but the final results agree with each other.}
\begin{ruledtabular}
\begin{tabular}{lcclc}
 RCC  & \multicolumn{2}{c}{RCC results} & RNCC  & RNCC   \\
 \cline{2-3}\\
  term        &  This work & Ref. \cite{sidh}   &  term   & result  \\
\hline
 & & & \\
$DT_1^{(1)}$                &  27.423    & 30.728   &  $DT_1^{(1)}$                  & 27.423 \\
$T_1^{(1)\dagger} D$        &  27.423    & 30.728   &  $\Lambda_1^{(1)} D$           & 21.837 \\
$T_1^{(0)\dagger}DT_1^{(1)}$&  $-1.756$  & $-1.554$ &  $\Lambda_1^{(0)} DT_1^{(1)}$  & $-0.715$ \\
$T_1^{(1)\dagger}DT_1^{(0)}$&  $-1.756$  & $-1.554$ &  $\Lambda_1^{(1)} DT_1^{(0)}$  & $-1.377$ \\
$T_2^{(0)\dagger}DT_1^{(1)}$&  $-3.594$  & $-1.564$ &  $\Lambda_2^{(0)} DT_1^{(1)}$  & 0.0 \\
$T_1^{(1)\dagger}DT_2^{(0)}$&  $-3.594$  & $-1.564$ &  $\Lambda_1^{(1)} DT_2^{(0)}$  & $-2.867$ \\
$T_1^{(0)\dagger}DT_2^{(1)}$&  0.112     & 0.121    &  $\Lambda_1^{(0)} DT_2^{(1)}$  & 0.036 \\
$T_2^{(1)\dagger}DT_1^{(0)}$&  0.112     & 0.121    &  $\Lambda_2^{(1)} DT_1^{(0)}$  & 0.0 \\
$T_2^{(0)\dagger}DT_2^{(1)}$&  1.008     & 1.030    &  $\Lambda_2^{(0)} DT_2^{(1)}$  & 0.950 \\
$T_2^{(1)\dagger}DT_2^{(0)}$&  1.008     & 1.030    &  $\Lambda_2^{(1)} DT_2^{(0)}$  & 0.981 \\
Others                      & $-0.892$   &  0.04  & Others                         & $-1.464$ \\
Normalization               &            &  $-7.717$        &                                &     \\
\end{tabular}
\end{ruledtabular}
\label{tab3}
\end{table}

We also compare contributions from different RCC terms (contributions from the h.c. terms are given separately) given in Ref. \cite{sidh} and from the present work in Table \ref{tab3}.
We quote explicitly contribution due to normalization of the wave function for the result reported in Ref. \cite{sidh} by multiplying the factor 1.157 listed in that reference. As can be 
seen normalization contribution is about 15\% in the PRCC method, which is absent in our result. Moreover, term-wise contributions also differ in both the works. Therefore, the results 
between both the works differ not only due to the inclusion 
of the contribution from the normalization of the wave function, but also due to different amplitudes of the RCC operators. In the above table, we also compare contributions 
from the RCC and RNCC terms to understand how the amplitudes in the RNCC method are changed from the RCC method. As can be seen contributions from the counter terms that replace 
h.c. terms of the CCSD method in the NCCSD method are significantly different. However, the final CCSD and NCCSD values are found to be very close. 
This supports validity of our results from our RCC methods. In addition, close agreement between the results from the CCSD(T) and CCSDTQ methods in the 
perturbed RCC theory and finite-field approach, respectively, justifies our claim for the high accuracy $\alpha_d$ calculations using these methods.

\section{Summary}\label{sec-sum}

We have carried out calculations of $\alpha_d$ of the Cd atom in the finite-field and perturbed RCC approaches. All-order RCC theory is employed at 
various levels of approximations to ascertain its accuracy.  We find our calculation is in good agreement with the previous theoretical results that are 
obtained by the quasi-relativistic and two-component relativistic calculations, but differ substantially from another calculation reported recently using 
a perturbed RCC approach similar to ours. Based on our analysis, we recommend the value 46.02(50) $ea_0^3$ rather than the the available experimental 
result $49.65 \pm 1.49 \pm 0.16 \ e a_0^3$.  This calls for performing further measurements of $\alpha_d$ of the above atom to verify our claim. We also 
observe that the correlation trends for the finite-field and the perturbed RCC approaches are different.

\section*{Acknowledgements}

We thank Professor U. Hohm for the personal communication and providing information on the published preliminary dipole polarizability value of Cd atom. 
We are grateful to Professor R. J. Bartlett and Dr. Ajith Perera for many useful discussions. B.K.S. acknowledges financial supports from Chinese
Academy of Science (CAS) through the PIFI fellowship under the project number 2017VMB0023 and TDP project of Physical Research Laboratory (PRL). Y.Y. is supported by the National Natural Science Foundation of
China under Grant No. 91536106, the CAS XDB21030300, and the NKRD Program of China (2016YFA0302104). Computations were carried out using Vikram-100 HPC cluster of PRL, Ahmedabad, India and HPC facility at 
Institute of Physics (IOP), CAS, Beijing, China.

\end{document}